\documentclass[twocolumn,pbl,aps]{revtex4}

 \usepackage[dvips]{graphicx}
 \usepackage[dvips]{graphics}

\begin{document}
\title{Comment on "No Evidence for Orbital Loop Currents in Charge Ordered ${\rm YBa_2Cu_3O_{6+x}}$ from
Polarized Neutron Diffraction"}

\author{P. Bourges}
\email{philippe.bourges@cea.fr} 
\affiliation{Laboratoire L{\'e}on Brillouin, CEA-CNRS, Universit\'e Paris-Saclay, CEA Saclay, 91191 Gif-sur-Yvette, France}

\author{Y. Sidis}
\email{yvan.sidis@cea.fr}
\affiliation{Laboratoire L{\'e}on Brillouin, CEA-CNRS, Universit\'e Paris-Saclay, CEA Saclay, 91191 Gif-sur-Yvette, France}

\author{L. Mangin-Thro}
\affiliation{Institut Laue-Langevin, 71 avenue des martyrs, 38000 Grenoble, France}

\date{\today}

% \pacs{PACS numbers: 74.25.Ha  74.72.Bk, 25.40.Fq }
%  25.40.Fq Inelastic neutron scattering
% 74.72.Bk Superconducting materials Y-based cuprates
% 74.25.Ha Superconductivity Magnetic properties

\begin{abstract}

Intra-unit cell magnetic order has been observed in four different families of high-temperature superconductors from polarized neutron diffraction experiments and supported by several other techniques. That order, which does not break translation symmetry, is consistent with the predicted  orbital moments generated by two microscopic loop currrents in each CuO$_2$ cell.  Recently, using polarized neutron diffraction, Croft {\it et al} [Phys. Rev. B 96, 214504 (2017)] claim to find no evidence for such orbital loop currents in charge ordered ${\rm YBa_2Cu_3O_{6+x}}$. Their experiment is done with detwinned samples at least 100 times smaller than in previous experiments without counting much longer. We show by a detailed quantitative analysis of their data that contrary to their conclusion, the  magnetic signal falls below their threshold of detection. None of the data reported by  Croft {\it et al} challenge the universality of the 
intra-unit cell order in cuprates.

\end{abstract}

\maketitle

\section{\label{Intro} Introduction}

In an extensive series of papers \cite{Fauque,Mook,Baledent-YBCO,Lucile15,Lucile17,Li2008,Li2011,Baledent-LSCO,DeAlmeida-Didry2012,Lucile14}, we and our collaborators  demonstrated using polarized neutron diffraction (PND)\cite{instruments} that the  pseudogap state of underdoped cuprate superconductors is characterized by a Q=0 magnetic order, also referred to as an intra unit cell (IUC) magnetic order\cite{CC-review,CC-review2}. That encompasses  results in four different cuprates families with a large variety of hole doping (p): $\rm YBa_2Cu_3O_{6+x}$  (YBCO) \cite{Fauque,Mook,Baledent-YBCO,Lucile15,Lucile17},  $\rm HgBa_2CuO_{4+\delta}$ \cite{Li2008,Li2011},   $\rm La_{2-x}Sr_xCuO_4$ \cite{Baledent-LSCO}  and  $\rm Bi_2Sr_2CaCuO_{8+\delta}$ \cite{DeAlmeida-Didry2012,Lucile14}.  Two reviews were written to give more experimental and technical details and put the different neutron results  in perspective with the other physical properties of high-temperature cuprates \cite{CC-review,CC-review2}. 

The PND  experiment we discuss here is very challenging\cite{CC-review}. The earlier results \cite{Fauque} revealed for the first time the IUC magnetic signal in five different YBCO samples and its stricking evolution with hole doping, following the pseudogap physics. Gradually, the data analysis was more quantitative in the subsequent publications in YBCO \cite{Mook,Baledent-YBCO,Lucile15,Lucile17}.  Over the years, the data analysis has been improved for quantitative accuracy (see for instance, the refined analysis on the sample C in Mangin-Thro  {\it et al} \cite{Lucile17}  compare to our original report in Fauqu\'e  {\it et al}\cite{Fauque}). This has important consequences on the magnetic signal amplitude and the confidence one can get from it. 

In a recent paper, Croft {\it et al}\cite{Hayden} claim that they found  no evidence for the appearance of magnetic order below 300 K in two   YBCO  samples. We show that  they could not observe the magnetic signal owing to the insufficient detection capability of their measurements (see \cite{comment-v2} for details).

$\bullet$ First, the neutron intensity is proportional to sample mass. By using samples at least $\sim$ 100 times smaller than ours on a spectrometer  with about 3 times larger neutron flux (at the used wavelength),  Croft {\it et al}\cite{Hayden} face about 30 times more experimental limitations. The counting times in previous reports\cite{Fauque,Mook,Baledent-YBCO,Lucile15,Lucile17,Li2008,Li2011,Baledent-LSCO,DeAlmeida-Didry2012,Lucile14} could reach 2 hours/point. Even after a counting time of 4 hours/point in some of their data, their experiment does not reach the required accuracy.  

$\bullet$ Second, Croft {\it et al}\cite{Hayden} erroneously overestimate by a factor $\sim$ 3  the magnetic signal that Fauqu\'e et al\cite{Fauque} have previously reported. This seems to be related to multiple simplifications of their analysis,  spanning incorrect data calibration,  misleading sample comparison,  ignorance of the impact of detwinning. 
Additional experimental limitations have been overlooked in \cite{Hayden}. Indeed, not determining the spin-flip reference line properly, not doing a polarization analysis and inadequate control of the flipping ratio of the neutron beam add to uncertainties in their measurements \cite{comment-v2}. 

$\bullet$ Third,  the comparison with local probes results in \cite{Hayden} is outdated and partial as it dismisses the recent literature about muon spin resonance results \cite{LeiShu,Pal17}. Recent muon spin rotation measurements report magnetic correlations 
at T* with finite time-scales ($\sim$ 10 ns),  which are fluctuating slowly enough to  appear static to neutrons. This observation is in disagrement with the conclusion  of Croft {\it et al}\cite{Hayden}, but fully consistent with our observation of IUC order \cite{Fauque,Mook,Baledent-YBCO,Lucile15,Lucile17,Li2008,Li2011,Baledent-LSCO,DeAlmeida-Didry2012,Lucile14}.  Finite time scale magnetic correlation can be associated with slowly fluctuating magnetic domains, such short range correlations have been actually reported in nearly optimally doped YBCO\cite{Lucile15} using PND.  This would have the effect to reduce the magnetic signal  by another factor of $\sim$ 3 in the experiment of  Croft {\it et al} \cite{Hayden}.

The claimed upper bound for a possible magnetic moment is therefore not correct and should be disregarded. None of the data reported by  Croft {\it et al} \cite{Hayden} disprove that the IUC magnetic order is universal in all cuprates. 

\section{\label{compare} Raw data comparison}

\begin{figure}[t]
\includegraphics[width=8cm,angle=0]{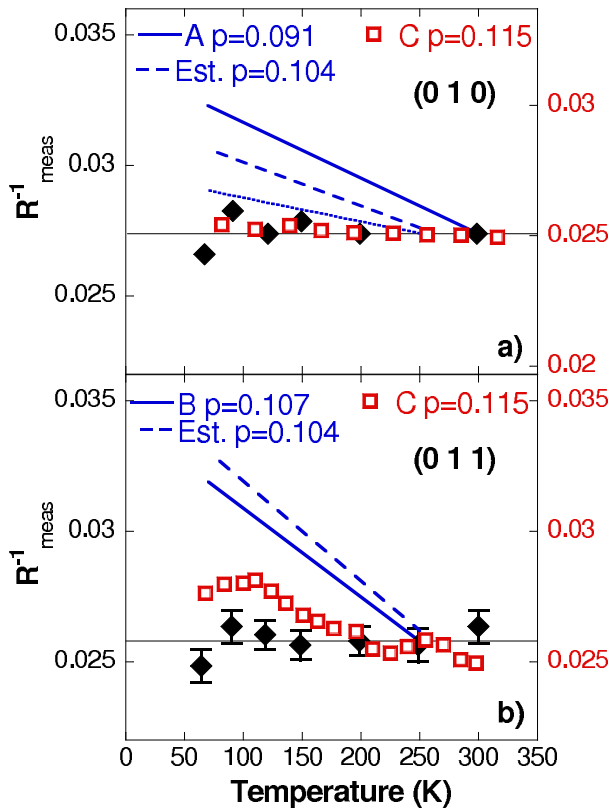}
\caption {
(color online)  Inverse of the measured flipping ratio,  $R^{-1}_{\rm meas}$, for two detwinned underdoped YBCO samples at two Bragg reflections (top panel) Q=(010) and (lower panel)  Q=(011). The black full symbols corresponds to the data of  Croft {\it et al}\cite{Hayden} (sample YBCO$_{6.54}$  with p=0.104) and the open red squares ones are those for sample C  YBCO$_{6.6}$ (p=0.115), of Fauqu\'e et al(ref. \cite{Lucile17} for the (010) reflection and refs. \cite{Fauque,CC-review}  for (011)). For the open symbols, the error bar is lower than the size of the point. All blue lines correspond to incorrect estimates reported by Croft {\it et al}\cite{Hayden} of the data of Fauqu\'e et al\cite{Fauque} (see \cite{comment-v2} for details).}
\label{invR}
\end{figure}

First, we compare the raw experimental data obtained by  Fauqu\'e {\it et al}\cite{Fauque,Lucile17} and Croft {\it et al}\cite{Hayden}. In  \cite{Fauque,Lucile17}, we have studied an array of co-aligned high-quality twin-free  single crystals. As the samples studied by Croft {\it et al}\cite{Hayden}, each single crystal was synthesized using the same self-flux method using a BaZrO$_3$ crucible as described in previous reports \cite{hinkov}  and  on which charge density wave order has been observed as well\cite{blanco-canosa}. 

As the IUC magnetic order does not break the symmetry of the lattice, one should study the inverse flipping ratio $R^{-1}_{\rm meas}$ at a Bragg position where the signal is expected to highlight a possible magnetic intensity at low temperature.  The inverse flipping ratio $R^{-1}_{\rm meas}$ 
reads:
%-------------------------------------------------------------
\begin{equation}
R^{-1}_{\rm meas}=\frac{I_{SF}}{I_{NSF}}= \frac{I_{mag}}{I_{NSF}}+ R^{-1}_{0}
\label{Imag}
\end{equation}
%-------------------------------------------------------------
where $I_{NSF}$ and $I_{SF}$ stand for the non-spin-flip and spin-flip intensities, respectively. As regularly emphasized\cite{Baledent-YBCO,Lucile15,Lucile17,DeAlmeida-Didry2012,Lucile14,CC-review,Hayden}, $R^{-1}_{\rm meas}$  is essentially a ratio of measured quantities and does not depend on any assumptions on nuclear structure factor or flipping ratio.  Changes in  $R^{-1}_{\rm meas}$ for the same Bragg peak should be comparable among the different studies. The second part of Eq. \ref{Imag} shows how  the magnetic intensity $I_{mag}$ can be extracted by comparison with the bare inverse flipping ratio $R^{-1}_{0}$ (background baseline) as it has been shown in several reports \cite{Baledent-YBCO,Lucile15,Lucile17,DeAlmeida-Didry2012,Lucile14}.  

We report in Fig. \ref{invR} the raw $R^{-1}_{\rm meas}$ at a Bragg positions (010) and (011) for two detwinned samples:  sample H1 with p=0.104 of Croft {\it et al}\cite{Hayden} and sample C of Fauqu\'e {\it et al}\cite{Fauque,Lucile17} (see \cite{comment-v2} for samples description). Within error bars, there is no disagreement between both data for the (010) reflection (Fig \ref{invR}.a).  However, Fig \ref{invR}.b shows a certain difference between both datasets for the (011) reflection whose possible origins can be understood as discussed below in the next  section (see also \cite{comment-v2}). For both reflections, a more surprising discrepancy occurs between the actual data of Fauqu\'e {\it et al}\cite{Fauque} and the alleged ones estimated by Croft {\it et al}\cite{Hayden}  (the dashed and dotted lines in their figures 8d and 8e). Obviously,  both quantities should exactly match but they do not. That underlines the erroneous analysis performed by Croft {\it et al}\cite{Hayden}, and this for both Bragg peaks. 
 We give in the supplemental material\cite{comment-v2} a discussion on possible origins of this discrepancy. 

Before showing alternative analyses of Croft {\it et al}\cite{Hayden} data, two remarks are necessary: 

$\bullet$  First,  the ratio  $\Delta R^{-1}_{mag}={I_{mag}}/{I_{NSF}}$ in Eq. \ref{Imag} does not change appreciably versus doping for the Bragg peak (011) in the doping range of interest here (p=0.1-0.12). $I_{NSF}$ corresponds to the nuclear structure factor $|F_N|^2$ where $F_N$ is given for instance by Eq. (12) in \cite{Hayden}. A simple calculation shows that nuclear structure factor for the Bragg reflection (011) versus oxygen content are decreasing  with increasing oxygen content similarly than the IUC magnetic intensity\cite{comment-v2}. Therefore, it cannot be objected that the ratio  $\Delta R^{-1}_{mag}$ should decrease upon doping as does the magnetic intensity like it is considered in \cite{Hayden}. As a result, similar $\Delta R^{-1}_{mag}$ is expected for both samples in Fig. \ref{invR} within a  20\% difference.

$\bullet$   Second, it should be noticed that the data shown for the Bragg peak (011) of sample C is the best example of a IUC magnetic signal ever reported in YBCO\cite{CC-review}. That corresponds to the highest experimentally reported ratio of the magnetic intensity compared to the nuclear intensity in Eq. \ref{Imag}:  $\Delta R^{-1}_{mag}$=0.25 \% at 70K for the (011) reflection\cite{CC-review}. All other reports in twinned samples are lower \cite{Fauque,Mook,Baledent-YBCO,Lucile15}; this is due to a larger nuclear intensity of the (101) Bragg intensity which is averaged with the (011) peak in twinned samples.

 Based on comparison of raw data in Fig. \ref{invR}, we demonstrate that the analysis of Croft {\it et al}\cite{Hayden} is highly questionable.  As shown in Fig. \ref{invR},  Croft {\it et al}\cite{Hayden}  allegedly estimate $\Delta R^{-1}_{mag}$=0.75 \% (see Fig. \ref{invR}) at odds with our results.

\begin{figure}[t]
\includegraphics[width=8cm,angle=0]{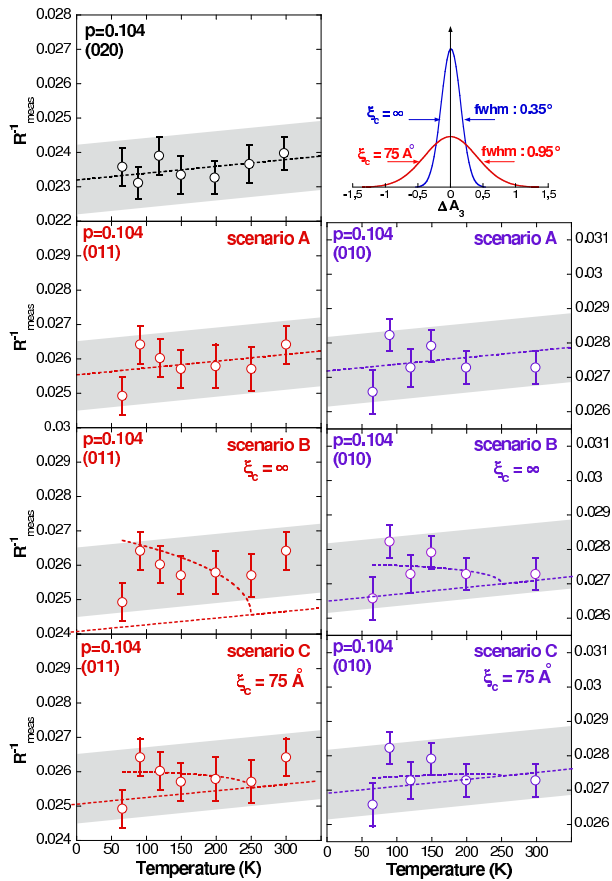}
\caption {
(color online)  1/R=$I_{SF}/I_{NSF}$ data for sample H1 (p=0.104) reproduced from Croft {\it et al} \cite{Hayden}.  The data points are exactly the same as those reported in Fig. 8 in \cite{Hayden} (reproduced as well in Fig. \ref{invR}) but  here we zoom on the data. The dashed lines are calculated for the three different scenarios considered in the text.    The top right panel represents A3 rocking scans convoluted  by the resolution function (blue) for the scenario B (red) for the scenario C.
}
\label{compareH1}
\end{figure}

\section{\label{improved} Alternative data analysis}

In \cite{comment-v2}, we show various points which can explain the discrepancy of Fig. \ref{invR} between the raw data of Fauqu\'e {\it et al} \cite{Fauque} and the alledged estimate made by Croft {\it et al} \cite{Hayden}.  For the Bragg peak (011),  the main error in  \cite{Hayden} comes from improper data calibration.  For the (010) Bragg position, the magnetic structure in twin-free sample  should be properly acknowledged. As a result, the full and dashed lines in the figures 8,9 and 11, of Croft {\it et al} \cite{Hayden} are all erroneous. Another global overestimated factor of 20\% has been neglected. Finally, a close comparison \cite{comment-v2} of the data analysis of both measurements reveal numerous limitations with the PND experiments  in \cite{Hayden}. 

Next, the dynamical nature of IUC magnetic signal as shown by recent muon spin resonance data \cite{LeiShu,Pal17}  have been ignored in \cite{Hayden} (see \cite{comment-v2} for details). Such low frequency fluctuations are necessarily related to the formation of finite-size magnetic domains \cite{Varma14}, corresponding to finite correlation lengths of the magnetic order observed in neutron diffraction. Such a short range correlation length of the IUC  order has been observed \cite{Lucile15} near optimally doped YBCO sample. In underdoped samples  (for dopings p$\sim$0.1-0.11), one can simply give an upper limit of the correlation length  along the c axis  $\xi_c \ge 75$ \AA \cite{Mook} as the observed magnetic peak is resolution limited due to large crystals mosaicity of $\sim 1^\circ$ in \cite{Mook}. In contrast, the measurements carried out by  Croft {\it et al} \cite{Hayden} are  performed on tiny samples with mosaic spread at least 10 times smalller. As shown in the top-right panel of Fig. \ref{compareH1},  a magnetic signal with a correlation length $\xi_c \sim 75$ \AA\ would exhibit a significantly broader rocking scan (A3) peak than for a long range magnetic order assumed in \cite{Hayden}.  The magnetic amplitude at the Bragg position would be reduced accordingly  $\sim$3 times compared to the one of true long range ordered state. Clearly, this factor 3 can be invoked to explain the difference  in Fig. \ref{invR} between the data of Fauqu\'e {\it et al}\cite{Fauque} and those of Croft {\it et al}\cite{Hayden} due to the different mosaicity of the samples.

Further, we turn to a consistent comparison of the magnetic signal reported by Fauqu\'e {\it et al}\cite{Fauque} to the experimental data of Croft {\it et al} \cite{Hayden}. In Fig.  \ref{compareH1}, we zoom   $R^{-1}_{\rm meas}=I_{SF}/I_{NSF}$ data for the sample p=0.104 of Croft {\it et al} \cite{Hayden}. The same treatment can be found in \cite{comment-v2} for the other sample (p=0.123) of Croft {\it et al} with the same conclusion. 

The inverse of flipping ratio at the reflection (020) is linear in temperature with a positive slope as it has been shown to exist \cite{Mook,Lucile17}. As discussed in \cite{comment-v2}, this slope is inevitable as the sample drifts in the neutron beam upon changing temperature.  Croft {\it et al} \cite{Hayden} arbitrarily describe it with a flat horizontal line only. At the accuracy required to observe the IUC  magnetic signal, this is not a correct approximation.  With the large  grey shaded area, we next represent the zone of the limit of detection,  corresponding to $\delta R^{-1} \simeq \pm 0.001$  \cite{comment-v2}, on both sides of an average sloping line. This area is due to combined effects of the statistical errors of each points, occurence of off-statistical points (possibly related to mechanical errors in  positioning) and the scarce number of points (especially above T*). This area is simply deduced from the measurements at the Bragg (020) reflection where no magnetic signal is expected for the IUC order.

 For clarity, the same error of $\delta R^{-1} \simeq \pm$ 0.1\% is used for all plots. That error is  typically  equivalent to an error of $\simeq \pm 2$ on the flipping ratio or $\delta I_{SF} \simeq \pm 4\%$ of the spin-flip intensity. This uncertainty is not negligible  as our best report of a magnetic signal is $\Delta R^{-1}_{mag} \simeq$ 0.25\% (Fig.  \ref{invR}). The error on the spin-flip intensity in previous PND experiments \cite{Fauque,Mook,Lucile17} leads to an estimate  of  $\delta R^{-1} \simeq \pm $ 0.01-0.02 \% (see  {\it e.g.} Fig. S1 in the Supplemental materials of Mangin-Thro {\it et al} \cite{Lucile17} for an example of measured error bars).

In Fig. \ref{compareH1}, the detection limit of $\delta R^{-1} \simeq \pm 0.1 \%$ is next reported for the Bragg reflections (010) and (011) where the magnetic signal is expected.  To compare that set of data, with those of Fauqu\'e {\it et al}\cite{Fauque,Lucile17}, we then consider three different scenarios which are plotted in Figs. \ref{compareH1}: A) no magnetic signal is present, B) a long range magnetic order is present  C) a magnetic order with short range correlation along the {\bf c}-direction is present (with the same integrated intensity of scenario (B)).  The amplitude in scenario (B) corresponds to the amplitude expected for a detwinned sample with the appropriate doping level\cite{comment-v2}. That basically corresponds to our best experimental evidence of a magnetic signal $\Delta R^{-1}_{mag} \simeq$ 0.25\% as plotted in Fig. \ref{invR}. The top right panel of Fig. \ref{compareH1} simulates the A3 rocking scan for the scenario B in blue and the scenario C in red where both hypothetical curves have been  convoluted by the resolution function given by the Fig. 6.a in \cite{Hayden}.

Following the various points discussed above and in line with the Fig. \ref{invR}, one clearly sees that the expected signal from  \cite{Fauque,Lucile15} is only marginally larger than the experimental uncertainty (error area). For all the three scenarios presented above, only one parameter, the overall level of the background of the baseline,  is fitted. All the other parameters are given from the literature \cite{Fauque} and the discussion above. Clearly, for all cases, scenarios A (of Croft {\it et al} \cite{Hayden}) and C (a short range along ${\bf  c^*}$ IUC magnetic order compatible with the report of ref. \cite{Mook}) cannot be distinguished at all. Even for a true long range magnetic order along ${\bf  c^*}$   (scenario B), the data are insufficient to eliminate with confidence the existence of IUC order due to indetermination of the baseline.  Clearly,
the sensitivity of the experiment is insufficient to observe the IUC order contrary to the claim of  Croft {\it et al} \cite{Hayden}. 

\section{\label{conclusion} Conclusion}

In conclusion,  Croft {\it et al} \cite{Hayden} do not have the experimental accuracy to observe the IUC signal that we have been reporting for the last decade in YBCO  \cite{Fauque,Mook,Baledent-YBCO,Lucile15,Lucile17}.  Different factors applied: first, the accuracy limit of the experiment   of Croft {\it et al} \cite{Hayden}  is represented by the shaded areas which indicate the uncertainty of  $\delta R^{-1}_{mag} \simeq \pm$ 0.1\% on the thermal dependence of the baseline for the ratio  $R^{-1}_0$ in Eq. \ref{Imag}. Second, the purported level of intensity of Fauqu\'e {\it et al} has been erroneously and systematically overestimated by a factor $\sim$ 3 (see Fig. \ref{invR}).  That corresponds to the scenario B of figs.  \ref{compareH1}, where a long range IUC order is assumed. Third, an additional factor 3 occurs if the signal is short ranged along L with $\xi_c\simeq 75$ \AA\   (scenario C of Figs.  \ref{compareH1}).  The most plausible scenario for the magnetic intensity lies in between scenarios B and C in Fig. \ref{compareH1}. This is clearly  below the detection limit of the data of Croft {\it et al} \cite{Hayden}. The claimed upper bound for a possible magnetic moment in  \cite{Hayden} is therefore baseless and should be disregarded. 

The IUC magnetic signal has been well documented for the last decade using polarized neutron diffraction\cite{Fauque,Mook,Baledent-YBCO,Lucile15,Lucile17,Li2008,Li2011,Baledent-LSCO,DeAlmeida-Didry2012,Lucile14}.  The magnetic signal, observed in four cuprate familiees, is seen only at specific Bragg positions. Data in YBCO and in $\rm HgBa_2CuO_{4+\delta}$ are nearly indistinguishable with a systematic doping dependence. In all cuprates, important polarization analysis has been conducted and the polarization sum rule, which demonstrates the magnetic nature of the signal, always nicely obeyed. 

The experiment is highly non-trivial with many technical pitfalls to miss the genuine signal. Among other features, it requires sufficient data at high temperature for a proper knowledge of the background. The IUC magnetic signal falls below the experimental sensitivity of the experiment of Croft {\it et al} \cite{Hayden} predominantly because too tiny YBCO samples were studied (at least 100 times smaller than in previous reports).  Further,  a large number of flaws and inaccuracies in their data analysis invalidates their comparison with the previous data\cite{comment-v2}. Their estimate of the magnetic signal previously reported could be incorrect up to an order of magnitude. None of the data shown invalidate that the IUC magnetic order is an intrinsic property of the pseudogap state of cuprates. 

{\bf acknowlegments}

 We wish to thank Gabriel Aeppli, Victor Bal\'edent, Dalila Bounoua, Johan Chang, Niels-Bech Christensen, Benoit Fauqu\'e, Martin Greven, Jaehong Jeong, Steve Kivelson, Yuan Li, Tasutomo Uemura and Chandra Varma for stimulating discussions on different aspects related to this work. We acknowledge financial supports from the project NirvAna (contract ANR-14-OHRI-0010) of the ANR French agency.


\begin{thebibliography}{100}
\bibitem{instruments} The measurements have been performed using various polarized neutron instruments:   triple-axis spectrometers, 4F1 at Laboratoire L\'eon Brillouin and IN22 Institut Laue Langevin (ILL), as well as a diffractometer D7 at ILL.
\end{thebibliography}
\end{document}